\documentclass[11pt,oneside,letterpaper]{article}
\usepackage{amssymb}
\usepackage{amsmath}
\usepackage[dvips]{graphicx}
\usepackage{setspace}
\usepackage{fancyhdr}
\usepackage{xcolor}
\usepackage{ifpdf}
\usepackage{graphicx}
\usepackage{rotating}
\usepackage{comment}

\def\p{\partial}

\def\N{\mathcal{N}}

\newcommand{\bi}{\begin{itemize}}
\newcommand{\ei}{\end{itemize}}
\newcommand{\bea}{\begin{eqnarray}}
\newcommand{\eea}{\end{eqnarray}}
\newcommand{\be}{\begin{equation}}
\newcommand{\ee}{\end{equation}}

\newcommand{\nuo}{{r}_++{r}_--{1\over\nu}\sqrt{{r}_+{r}_-(\nu^2+3)}}

\numberwithin{equation}{section}
\allowdisplaybreaks  % allow page breaks in displayed eqs

\begin{document}

\vspace*{2.5cm}
\begin{center}
{ \LARGE \textbf{Warped AdS$_3$  Black Holes }\\}
\vspace*{1.7cm}
Dionysios Anninos, Wei Li, Megha Padi, Wei Song* and Andrew Strominger
\vspace*{0.6cm}
\newline
Center for the Fundamental Laws of Nature\\
Jefferson Physical Laboratory, Harvard University, Cambridge, MA, USA\\

\vspace*{0.8cm}

\end{center}
\vspace*{1.5cm}
\begin{abstract}
\noindent
Three dimensional topologically massive gravity (TMG) with a negative cosmological constant $-\ell^{-2}$ and positive Newton constant $G$ admits an AdS$_3$ vacuum solution for any value of the graviton mass $\mu$. These are all known
to be perturbatively unstable except at the recently explored chiral point $\mu\ell=1$. However we show herein that for every value of $\mu\ell\neq 3$ there are two other (potentially stable) vacuum solutions given by $SL(2,\mathbb{R})\times U(1)$-invariant warped AdS$_3$ geometries, with a timelike or spacelike $U(1)$ isometry.
 Critical behavior occurs at $\mu\ell=3$, where the warping transitions from a stretching to a squashing, and there are a pair of warped solutions with a null $U(1)$ isometry.  For $\mu\ell>3$, there are known warped black hole solutions which are asymptotic to warped AdS$_3$. We show that these black holes are discrete quotients of warped AdS$_3$ just as BTZ black holes are discrete quotients of ordinary AdS$_3$. Moreover new solutions of this type, relevant to any theory with warped AdS$_3$ solutions, are exhibited.   Finally we note that the black hole thermodynamics is consistent with the hypothesis that, for $\mu\ell > 3$, the warped AdS$_3$ ground state of TMG is holographically dual to a 2D boundary CFT with central charges $c_R =  {15(\mu\ell)^2+81\over G\mu((\mu\ell)^2+27)}$ and $c_L ={12 \mu\ell^2\over G((\mu\ell)^2+27)}$.

%We squash the living bejesus out of AdS$_3$ and find black holes with the living bejesus squashed out of them.

\end{abstract}

\vspace{1.0cm} *On leave from the Institute of Theoretical Physics,
Academia Sinica, Beijing,  China
\newpage
\setcounter{page}{1}
\pagenumbering{arabic}

\tableofcontents
\setcounter{tocdepth}{2}

%\cfoot{\thepage}
\onehalfspacing

%\newpage
\section{Introduction and summary}
Topologically massive gravity \cite{Deser:1981wh,Deser:1982vy} is described by the action
\be
I_{TMG} = \frac{1}{16\pi G}\left[\int d^3x\sqrt{-g}(R+2/\ell^2)+{1 \over \mu}I_{CS}\right]
\ee
where $I_{CS}$ is the gravitational Chern-Simons action (given explicitly below) and we
take both  $G$ and $\mu$ positive\footnote{For a recent discussion of the negative $G$ case see \cite{Carlip:2008jk, Li:2008yz}.}. For every value of the coupling $\mu$ TMG has a classical AdS$_3$ solution with radius $\ell$.  For large $\ell$ (and positive $G$) the linearized excitations about AdS$_3$ describe a propagating graviton with positive mass $\mu$, but negative energy (as well as the usual massless positive-energy gravitons). Hence the AdS$_3$  vacua are generically expected to be unstable and the quantum theory appears ill-defined. However at the critical value $\mu\ell=1$ the Compton wavelength of the massive graviton reaches the AdS$_3$ radius and the linearized energy spectrum of asymptotically AdS$_3$ excitations is non-negative.\footnote{ By asymptotically AdS$_3$ we mean in the usual sense of Brown and Henneaux \cite{Brown:1986nw} or Fefferman and Graham \cite{fg}. It was verified  in \cite{Hotta:2008yq} that these boundary conditions are consistent for TMG in that the generators of the asymptotic symmetry group are well-defined finite expressions. It is however possible \cite{kw} that there is more than one set of consistent AdS$_3$ boundary conditions defining inequivalent theories. Recently an alternate definiton of the theory with logarithmically weaker boundary conditions, referred to as ``cosmological topologically massive gravity at the chiral point", has been proposed \cite{Grumiller:2008qz}, but  the asymptotic symmetry group or the finiteness of its generators have yet to be studied.  It was established  \cite{Grumiller:2008qz} that if the theory can be consistently defined in this manner it is unstable, does not have a hermitian Hamiltonian, and is potentially dual to a logarithmic CFT.}  Accordingly it was conjectured \cite{Li:2008dq} that a consistent  quantum theory  of so-called chiral gravity can be defined at $\mu\ell=1$.

  In this paper we
  shall focus on non-chiral values of $\mu\ell$. The fact that for $\mu\ell\neq 1$
  the AdS$_3$ vacua are all unstable does not preclude the possibility that these theories have other stable ground states around which they can be consistently expanded.  With this in mind we look for other vacua and find that there are in fact two warped AdS$_3$ vacua (some of which were known already in \cite{Vuorio:1985ta,Percacci:1986ja}) for every value of $\mu$. Ordinary AdS$_3$ can be viewed as a fibration of the real line over AdS$_2$. Warped AdS$_3$ is similar but with a constant warp factor multiplying the fiber metric. This breaks the $SL(2,\mathbb{R})_L\times SL(2,\mathbb{R})_R$ isometry group of AdS$_3$ down to $SL(2,\mathbb{R})\times U(1)$. The fiber is stretched (squashed) for $\mu\ell >3$ ($\mu \ell <3$) and there are solutions with both timelike and spacelike $U(1)$ isometries. At the critical value $\mu\ell=3$ there are still two warped solutions, but both have a null $U(1)$ isometry.

  The curvature of the warped  solutions is of order $\mu^2$. Hence the Compton wavelength of the massive graviton is always of order the radius of curvature and it cannot be decoupled from that of the background by taking  $\mu\to \infty$ as in the ordinary AdS$_3$ case. Therefore it is not a priori obvious  whether or not the massive gravitons lead to instabilities of the warped  vacua. The first (as yet untaken) steps are to understand the boundary conditions for warped AdS$_3$, and to solve for the linearized spectrum. The reduced isometry group makes these tasks substantially more difficult than for ordinary AdS$_3$.  At present we do not know whether or when the warped vacua are perturbatively stable.  This is a key issue and we hope to return to it at a later point.

  For $\mu\ell >3$, there are known regular black hole solutions \cite{Nutku:1993eb,Gurses,Bouchareb:2007yx} which are asymptotic to warped AdS$_3$ with a spacelike $U(1)$. We show herein that these
  warped black holes are discrete quotients by an element of  $SL(2,\mathbb{R})\times U(1)$ of warped AdS$_3$, just as BTZ black holes are discrete quotients of AdS$_3$ \cite{Banados:1992wn,Banados:1992gq,Coussaert:1994tu}. We further describe a fascinating zoo of other solutions of this type for other values of $\mu$.

  It is important to note that warped AdS$_3$ arises in a number of contexts besides TMG, see e.g. \cite{Gurses, Rooman:1998xf,Duff:1998cr, Israel:2003ry,Andrade:2005ur, Bengtsson:2005zj,Banados:2005da, Son:2008ye,Balasubramanian:2008dm}. It follows from our quotient construction that the warped black holes are solutions of all these theories as well.  In addition warped AdS$_3$ is
 a submanifold (at fixed polar angle) of the near horizon geometry of extremal Kerr \cite{Bardeen:1999px,Bengtsson:2005zj}. We will report elsewhere on the application of our results to Kerr \cite{ghs}.

  In AdS$_3$ the group of elements defining the quotient selects a left and right temperature $T_L$ and $T_R$ of the boundary CFT. Using $c_{L,R}={3\ell/2G}$ (and assuming unitarity), the density of states of the boundary CFT exactly matches the Bekenstein-Hawking entropy of the corresponding black hole \cite{Strominger:1997eq},  thereby explaining the latter. A similar exact match is found for warped black holes in warped AdS$_3$ provided the central charges are
  \be
  c_R =  {15(\mu\ell)^2+81\over G\mu((\mu\ell)^2+27)}\label{ccl}\ee \be c_L ={12 \mu\ell^2\over G((\mu\ell)^2+27)}.\label{ccr}\ee
  As this picture fits together rather nicely we cannot resist conjecturing that $\mu\ell>3$ TMG defined with suitable asymptotically warped boundary conditions exists and is dual to a 2D boundary CFT with central charges \ref{ccl} and \ref{ccr}.\footnote{We do not have a conjecture for $\mu\ell \le 3$ except of course for $\mu\ell=1$.} Some non-trivial - but far from definitive - evidence for the conjecture is given herein.

  In the next section we define the theory and our conventions. In section 3 we find the classical warped AdS$_3$ vacua for all $\mu$. In section 4 we review the known warped black hole solutions. In section 5 we show that they are quotients of warped AdS$_3$.
  Section 6 describes the general quotient solution.
  In section 7 we describe the black hole thermodynamics and formulate a conjecture on the existence of a boundary CFT dual to warped AdS$_3$.

 After this work was completed several works appeared \cite{cdww,gps} which overlap with section 3 as well as \cite{ Adams:2008wt,Herzog:2008wg,Maldacena:2008wh} which finds the null black holes of section 6.3 in the context of cold atoms.

\section{Framework}

Our story begins with the action for three dimensional topologically
massive gravity with a negative cosmological constant,
\begin{multline}
I_{TMG} = \frac{1}{16\pi G}\int_{\mathcal{M}}d^3x\sqrt{-g}\left(R+2/\ell^2\right)
\\+\frac{\ell}{96\pi G \nu}\int_{\mathcal{M}} d^3x\sqrt{-g}\varepsilon^
{\lambda\mu\nu}\Gamma^{r}_{\lambda \sigma}\left(\partial_{\mu}\Gamma^
\sigma_{{r} \nu}+\frac{2}{3}\Gamma^\sigma_{\mu\tau}\Gamma^\tau_{\nu
{r}} \right)
\end{multline}
where $\varepsilon^{\tau \sigma u}=+1/\sqrt{-g}$ is the Levi-Civita
tensor and $G$ has the conventional positive sign.  The coefficient
of the Chern-Simons action involves the dimensionless coupling $\nu$
which is related to the graviton mass $\mu$ by
\be \nu=\frac{\mu \ell}{3} \ee
For asymptotically AdS$_3$ spacetimes there is a critical chiral
gravity theory at $\mu \ell=1$ or $\nu =\frac{1}{3}$ \cite{Li:2008dq}. However our main interest in this paper are the warped AdS$_3
$ vacua. These exhibit critical behavior at  $\nu=1$ or $\mu\ell=3$. At the risk of confusion we use $\nu$ rather than $\mu$
as the formulae are significantly simpler.  Since the
Chern-Simons is parity odd without loss of generality we may take $\nu$
to be positive.

Upon varying the above action with respect to the metric we obtain
the bulk equation of motion,
\be
{G}_{\mu\nu}-\frac{1}{\ell^2}g_{\mu\nu}+{\frac{\ell}{3\nu}}C_{\mu\nu}=0
\label{eq:EOM}
\ee
where $G_{\mu \nu}$ is the Einstein tensor and $C_{\mu \nu}$ is
  the Cotton tensor:
\be
C_{\mu\nu} = \varepsilon_\mu^{~\alpha \beta}\nabla_\alpha\left(R_
{\beta\nu}-\frac{1}{4}g_{\beta\nu}R\right)
\ee
 From the equations of motion we see that any Einstein vacuum
solution is also a solution of TMG since $G_{\mu \nu}=g_{\mu\nu}/
\ell^2$. There are also non-Einstein solutions that solve these
equations of motion which we now proceed to explore.

\section{$SL(2,\mathbb{R})\times U(1)$-invariant TMG vacua}

The simplest  non-Einstein solution to TMG will be referred to as \emph
{warped AdS$_3$}, as it involves a warped fibration. For every value
of $\nu$ there are two such solutions, and the structure changes
qualitatively at $\nu=1$. The timelike warped case was previously
discovered as a solution of  TMG by Nutku and G\"urses
\cite{Nutku:1993eb,Gurses}. It has also been studied in various
other contexts
\cite{Rooman:1998xf,Duff:1998cr,Israel:2003ry,Israel:2004vv,Andrade:2005ur,Detournay:2005fz,Bengtsson:2005zj,Banados:2005da,Compere:2007in}.
To find the general solution we begin by recalling that AdS$_3$ can
be written as a kind of Hopf fibration over Lorentzian (or
Euclidean) AdS$_2$ where the fiber is the real line
\cite{Duff:1998cr,Strominger:1998yg,Bengtsson:2005zj}. The metric
can be written as a spacelike (or timelike) fibration with fiber
coordinate $u$ (or $\tau$)
\begin{eqnarray}
{ds^2} &=& \frac{\ell^2}{4}\left[-\cosh^2\sigma d\tau^2 + d\sigma^2 +
(du + \sinh\sigma d\tau)^2\right] \label{eq:eads}\\
  &=& \frac{\ell^2}{4}\left[ \cosh^2\sigma du^2 + d\sigma^2  - (d\tau
+ \sinh\sigma du)^2\right] \label{eq:hads}
\end{eqnarray}
with $\{u,\tau,\sigma\} \in [-\infty,\infty]$ and we have replaced $u$ with $-u$ in \ref{eq:hads}.
The $SL(2,\mathbb{R})_L\times SL(2,\mathbb{R})_R$ isometries are
given in appendix \ref{section:appA}. In order to obtain warped AdS$_3$
we must multiply the fiber by a warp factor, which breaks the
isometry group to $SL(2,\mathbb{R})\times U(1)$, where the $U(1)$ is
noncompact. Solutions of this type  fall into six different classes.
\subsection{Spacelike}
A \emph{spacelike} (or hyperbolic) warped anti-de Sitter solution is
given by warping \ref{eq:eads} in the form
\be
{ds^2} = \frac{\ell^2}{(\nu^2 + 3)}\left[  - \cosh^2\sigma d\tau^2 + d
\sigma^2 + \frac{4\nu^2}{\nu^2 + 3} \left(du + \sinh\sigma d\tau
\right)^2 \right]
\label{eq:ssqads}
\ee
For  $\nu^2 > 1$, the warp factor is greater than unity, so we have
spacelike \emph{stretched} AdS$_3$.
  If $\nu^2 < 1$ we have spacelike \emph{squashed} AdS$_3$. The isometries
preserved by \ref{eq:ssqads} are given by $U(1)_L\times SL(2,\mathbb{R})_R$. Note that the solution is not warped at $\nu=1$.

\subsection{Timelike}
A \emph{timelike} (or elliptic) warped anti-de Sitter solution is
given by a warping of \ref{eq:hads},
\be
{ds^2}= \frac{\ell^2}{(\nu^2  + 3)}\left[  \cosh^2\sigma du^2 + d
\sigma^2  - \frac{4\nu^2}{\nu^2 + 3} (d\tau + \sinh\sigma du)^2\right]
\label{eq:tsqads}
\ee
Once again if $\nu^2 > 1$ we have timelike stretched AdS$_3$. If $\nu^2 < 1$ we have timelike squashed AdS$_3$.
  As explored in \cite{Rooman:1998xf,Bengtsson:2005zj} timelike
stretched AdS$_3$ has closed timelike curves and is essentially  the
G\"{o}del spacetime.  While such spacetimes are quite interesting in
their own right,  we are limiting the scope of this paper to
spacetimes without naked CTCs.

We may also write the squashed metric in the global coordinates\footnote{These coordinates are OK locally but have global singularities for the stretched case, as do the spacelike warpings obtained by the double Wick rotation $t\to it,~~ \theta \to i \theta$}
\begin{multline}
\frac{ds^2}{\ell^2}= -dt^2+\frac{ d{r}^2}{{r}\bigl((\nu^2+3){r}+4\bigr)} +
2\nu {r}
dtd\theta
+\frac{{r}}{4}\left(3(1-\nu^2 ){r}+4\right)d\theta^2
\label{eq:ng}
\end{multline}
The origin is at ${r}=0$ where the $(t,\theta)$ metric determinant vanishes. Regularity requires the identification
\be
\theta \sim \theta+2\pi.
\ee
The coordinate transformation relating the two metrics is a special case of equations \ref{cxd}-\ref{eq:ctran} below.

\subsection{Null}
\emph{Null} (or parabolic) warped AdS$_3$s \cite
{Detournay:2005fz} are given by the metrics,
\be
{ds^2}={\ell^2} \left[ \frac{du^2}{u^2} +
\frac{dx^+ dx^-}{u^2} \pm  \left(\frac{dx^-}{u^2}
\right)^2 \right]
\label{eq:null}
\ee
with coordinate range $x^\pm \in [-\infty,\infty]$ and $u\in [0,
\infty]$. The above metrics are a solution to TMG only for $\nu^2 = 1$, and can be obtained as a kind of pp-wave limit in which $\nu \to 1$ of the $\nu \neq 1$ warped vacua. Pure AdS$_3$ is retrieved by dropping the last  $(dx^-)^2$ term.
The isometry group is $SL
(2,\mathbb {R})\times U(1)_{null}$ where the $U(1)_{null}$ is in the
null direction. Null warped AdS$_3$ has recently
appeared in a completely different context in the search for a dual
theory for cold atoms \cite{Son:2008ye,Balasubramanian:2008dm}. It
would be interesting to see whether the rest of the warped anti-de
Sitter metrics, or the null black hole solutions discussed below,  are related to this story.
\vspace*{0.5cm}

In summary, for  $\nu^2< 1$ there are timelike and spacelike squashed vacua, while for
$\nu^2>1$ there are timelike and spacelike stretched vacua. At $\nu=1$ there are two null warped vacua. All of these vacua have an  $SL(2,\mathbb{R})\times
U(1)$ isometry. In addition for every $\nu$ there is the usual
$SL(2,\mathbb{R})_L\times SL(2,\mathbb{R})_R$-invariant AdS$_3$ vacuum.

\section{Spacelike stretched black holes}

We now switch gears and study the black hole solutions which are
asymptotic to warped AdS$_3$.  Solutions which are free of naked
CTCs or other pathologies are until now known only for the
asymptotically spacelike stretched (i.e. $\nu^2>1$) case
\cite{Bouchareb:2007yx}.\footnote{These will be generalized below
for some other asymptotics.}  These will be reviewed in this
section. Closely related black objects were first discussed in
\cite{Nutku:1993eb,Gurses} and further studied in
\cite{Moussa:2003fc,Bouchareb:2007yx} where their conserved ADT
charges \cite{Abbott:1982jh,Deser:2002rt,Deser:2002jk,Deser:2003vh}
and some of their thermodynamic properties were computed.

The metric describing the spacelike stretched  black holes for $\nu^2>1$ is given in Schwarzschild coordinates by
\begin{multline}
\frac{ds^2}{\ell^2}=dt^2+\frac{ d{r}^2}{(\nu^2+3)({r}-{r}_{+})({r}-{r}_{-})} + \left(2\nu {r}  - \sqrt{{r}_{+}{r}_{-}(\nu^2+3)}\right)dtd\theta \\
+\frac{{r}}{4}\left(3(\nu^2 - 1){r}+(\nu^2+3)({r}_+ + {r}_-) - 4\nu\sqrt{{r}_+{r}_-(\nu^2+3)}\right)d\theta^2
\label{eq:ng}
\end{multline}
where ${r}\in[0,\infty]$, $t\in[-\infty,\infty]$ and $\theta \sim \theta + 2\pi$. In the ADM form the above metric becomes,
\be
ds^2=-N(r)^2 dt^2+\ell^2R({r})^2(d\theta+N^\theta(r) dt)^2+{\ell^4d{r}^2\over4
R({r})^2N(r)^2}
\ee
where we have defined,
\begin{eqnarray}
R({r})^2&\equiv&{{r}\over4}\left(3(\nu^2-1){r}+(\nu^2+3)({r}_++{r}_-)-4\nu\sqrt{{r}_+{r}_-(\nu^2+3)}\right)\\
N({r})^2&\equiv&{\ell^2(\nu^2+3)({r}-{r}_+)({r}-{r}_-)\over4 R({r})^2}\\
N^\theta({r})&\equiv&{2\nu{r}-\sqrt{{r}_+{r}_-(\nu^2+3)}\over2R({r})^2}
\end{eqnarray}
The horizons are located at ${r}_+$ and ${r}_-$ where $1/g_{{r}{r}}$ as well as the determinant of the $(t,\theta)$ metric vanishes. The vacuum solution for the black holes is given by ${r}_+={r}_-=0$ and, like $M=J=0$ BTZ, is singular at the origin ${r}=0$.
We also note that the above metric reduces to the metric of the BTZ black hole in a rotating frame when $\nu^2 = 1$. In the parameter region given by $\nu^2 > 1$ we have physical black holes so long as ${r}_+$ and ${r}_-$ are non-negative.
For $\nu^2 < 1$ we always encounter closed timelike curves at large values of ${r}$ when $\theta$ is identified. Such geometries will not be considered in this paper.

Finally we mention that the spacelike stretched black hole was studied in a slightly different coordinate system in \cite{Bouchareb:2007yx} and we give the details of this metric in appendix \ref{sec:bc}. At this point we have finished describing the basic geometry of the previously known non-Einstein black hole solutions of TMG.

\section{Warped black holes as quotients }
\label{sec:quot}

In this section we will show that the known warped black holes are quotients of warped AdS$_3$ under a discrete subgroup $\Gamma$ of the isometry group, much as BTZ black holes are quotients of AdS$_3$ \cite{Banados:1992wn,Banados:1992gq}.

 \subsection{Curvature invariants}

A first hint that warped black holes are locally equivalent to warped AdS$_3$ comes from looking at the coordinate invariant quantities. In three dimensions these are built from the Ricci tensor and its derivatives. It turns out that warped AdS$_3$ and the warped black hole solutions have the same values for these curvature invariants. The invariants built out of the Ricci tensor are given by,
\be
\{R, R_{\mu\nu}R^{\mu\nu}, R_{\mu\nu}R^{\mu\beta}R^{\nu}_{\ \beta}\}=  \frac{6}{\ell^2}\left\{-1,\frac{3 - 2\nu^2 +\nu^4}{\ell^2},\frac{-9+9\nu^2-3\nu^4-\nu^6}{\ell^4}\right\}
\ee
The derivatives of the Ricci tensor give rise to the following invariant,
\begin{equation}
\nabla_\mu R^{\beta \alpha}\nabla_\alpha R_\beta^\mu = \frac{18\nu^2(\nu^2-1)^2}{\ell^6}
\end{equation}
The above agreement between the coordinate invariant quantities of the warped AdS$_3$ and the warped black holes suggests that they are in fact locally equivalent (see \cite{Sousa} for a rigorous discussion of local equivalence in three dimensions) as we will now demonstrate directly.

\subsection{Coordinate transformations}

In this subsection we exhibit a local coordinate transformation from warped AdS$_3$ to the warped black hole metric and thus establish the claim that known warped black holes are locally equivalent to warped AdS$_3$. When we are in the parameter range $\nu^2 > 1$ the transformation between \ref{eq:ssqads} and \ref{eq:ng} is,
\begin{eqnarray}
\tau   &=& \tan^{-1}\left[\frac{2\sqrt{({r}-{r}_+)({r}-{r}_-)}}{2{r}-{r}_+ -
{r}_-}\sinh\left(\frac{1}{4}({r}_+-{r}_-)(\nu^2+3)\theta\right)\right]\label{cxd}
\end{eqnarray}
\begin{multline}
\quad \quad \text{  } u\quad = \quad \frac{\nu^2+3}{4\nu }\left[2t+\left(\nu({r}_++{r}_-)-\sqrt{{r}_+{r}_-(\nu^2+3)}\right) \theta\right] \\
- \tanh^{-1}\left[ \frac{{r}_+ + {r}_- - 2{r}}{{r}_+ - {r}_-}\coth\left(\frac{1}{4}({r}_+ - {r}_-)(\nu^2+3)\theta \right)\right] \label{eq:ident}
\end{multline}
\begin{eqnarray}
\sigma &=&   \sinh^{-1}\left[\frac{2\sqrt{({r}-{r}_+)({r}-{r}_-)}}{{r}_+ - {r}_-}\cosh\left(\frac{1}{4}({r}_+-{r}_-)(\nu^2+3)\theta \right)\right]
\label{eq:ctran}
\end{eqnarray}
A similar transformation takes us from \ref{eq:tsqads} to \ref{eq:ng} when we are in the parameter range $\nu^2  < 1$ although the solutions are not regular in that region. The above transformation breaks down for extremal black holes and we present instead the coordinate transformation in appendix \ref{section:appB}.

\subsection{Warped AdS$_3$ quotients}
Since the warped black holes are locally warped AdS$_3$, they must be a quotient of the latter by a discrete subgroup $\Gamma$ of the $SL(2,\mathbb R)\times U(1)$ isometries - in direct analogy with the case of the BTZ black hole \cite{Banados:1992gq}. In this section we find $\Gamma$. Let us identify points $\cal P$ in warped AdS$_3$ under the action of a Killing vector $\xi$ defining a one parameter subgroup of the full isometry group as follows \cite{Banados:1992gq},
\be
\mathcal {P} \sim e^{2\pi k \xi}\mathcal{P}, \quad \quad  k = 0, 1, 2 \dots
\ee
The isometries of the various types of warped AdS$_3$ are given in appendix \ref{section:appA}. From \ref{eq:ident} we see
that in order for the coordinate transformation to reproduce the black hole we must identify points along the $\partial_\theta$ direction such that $\theta \sim \theta + 2\pi$. Expressing the $\partial_\theta$  Killing vector in terms of the original warped anti-de Sitter coordinates, we discretely quotient along the isometry
\begin{equation}
2\pi \xi = \partial_\theta = {\nu^2+3\over8}\left[\left({r}_+ + {r}_- - {\sqrt{(\nu^2+3){r}_+{r}_-}\over\nu}\right)J_2-({r}_+ - {r}_-)\tilde{J}_2\right]
\label{eq:pphi}
\end{equation}
where the Killing vectors $J_2 \in U(1)_L$ and $\tilde{J}_2\in SL(2,\mathbb{R})_R$ are given in the appendix. Defining the left and right moving temperatures
\begin{eqnarray}
T_{R} &\equiv& {(\nu^2+3)({r}_+-{r}_-)\over8\pi\ell}\\
T_{L} &\equiv& {(\nu^2+3)\over8\pi\ell}\left({r}_+ + {r}_- -
{\sqrt{(\nu^2+3){r}_+{r}_-}\over\nu}\right)
\label{tltr}
\end{eqnarray}
we have
\be
\partial_{\theta} = \pi\ell(T_LJ_2-T_R\tilde{J}_2)
\ee
$T_L$ and $T_R$ are referred to as temperatures in analogy with the BTZ case where it is known \cite{Maldacena:1998bw} that the coefficients of the shifts are the temperatures of the
dual 2d CFT.
In the coordinates \ref{eq:ssqads}  the norm  is
\begin{eqnarray}
\nonumber |T_L J_2-T_R \tilde{J}_2|^2&=&{12\ell^2(\nu^2-1)\over(\nu^2+3)^2}\left[T_R\cos(\tau)\cosh(\sigma)\right]^2\\
\nonumber&+&{32\nu^2\ell^2\over(\nu^2+3)^2}T_LT_R\cos(\tau)\cosh(\sigma)\\
&+&{4\ell^2\over\nu^2+3}T_R^2+{16\nu^2\ell^2\over(\nu^2+3)^2}T_L^2
\end{eqnarray}
Note that for the squashed case the norm is negative at the boundary so a $\theta$ quotient would produce CTCs.

In summary the spacelike stretched black hole solution \ref{eq:ng} is a quotient of
spacelike stretched AdS$_3$ \ref{eq:ssqads} by a $2\pi$ shift generated by the isometry \ref{eq:pphi}.

\section{More general quotients}
In this section we consider more general quotients of warped AdS$_3$ by discrete subgroups of the isometry group which preserves at least $U(1)\times U(1)$. In appendix \ref{section:appA} the classification of such discrete subroups is extracted from \cite{Banados:1992gq}. There are a rich variety of possible structures and they differ for the six possible warpings as discussed below.

Of particular interest are quotients that correspond to regular black holes.  By this we mean a geometry with a geometric or causal singularity hidden by an event horizon. Geometric singularities arise when the discrete subgroup has a fixed point. Causal singularities arise when the identification produces null or timelike closed curves. In either case the norm of the quotient-generating  isometry has a zero. This singularity should be shielded from infinity by an event horizon, i.e. a place where the $N^2$ multiplying $dt^2$ in the ADM form of the metric vanishes.

We note that this might not be the only type of geometry which should be thought of as a black hole. For example the boundary of the Poincare patch of AdS$_2$ or AdS$_3$ is an event horizon when these spaces arise as near-horizon limits of extremal black holes or strings \cite{Spradlin:1999bn}. There is a Killing horizon associated to time translations but no singularity behind it. These spaces, though free of singularities, can be thought of as black holes. Also one may consider black hole geometries  - such as the black holes in a G\"{o}del universe - which have a regular region outside the horizon but CTCs at infinity. Hence the following discussion is a beginning and does not exhaust all interesting possibilities for black holes.

\subsection{Spacelike Warped}
We begin with the spacelike warped case. The three types of one-parameter subgroups are generated by
\begin{eqnarray}
\eta_a:&& \beta_2 J_2+\alpha_0 \tilde{J}_0\\
\eta_b:&& \beta_2J_2+\alpha_2 \tilde{J}_2\\
\eta_c:&& \beta_2 J_2+\tilde{J}_0+\tilde{J}_2
\end{eqnarray}
The norms of the Killing vectors generating the subgroups $\eta_a$, $\eta_b$ and $\eta_c$ are given in \ref{eq:snorm1}, \ref{eq:snorm2}, \ref{eq:snorm3}. The $\eta_c$ type identifications can be thought of as extremal limits of the previous two cases.
\subsubsection{Stretched}
We begin with the \emph{stretched} case. If we quotient along
$\eta_b$ with $\alpha_2 \neq 0$ we obtain the warped black hole
solution \ref{eq:ng}, as we already discussed. In order to locate
the horizon, we go to the coordinate system where $\partial_\phi =
\eta_b$, $\partial_t = J_2$ and $r =
-\frac{\alpha_2}{\beta_2}\cosh\sigma\cos\tau$. Then, $N^2$ in the
ADM decomposition is given by
\begin{equation}
N^2 =\frac{ (J_2 \cdot \eta_b )^2}{|\eta_b|^2} -|J_2|^2
\end{equation}
 The zeroes of $(N^2|\eta_b|^2)$ give the locations of the horizons, and the zeroes of $|\eta_b|^2$ determine the location of the causal singularities. Explicitly these are located at
\begin{equation}
r_\pm\equiv\pm|\frac{\alpha_2}{\beta_2}|,\quad\hbox{and},\quad
r_{s\pm}\equiv
\frac{-4\nu^2\pm\sqrt{(\nu^2+3)(4\nu^2-3(\nu^2-1)\frac{\alpha_2^2}{\beta_2^2})}}
{3(\nu^2-1)}
\end{equation}
where $r_\pm$ denote the horizons and $r_{s+}$ denotes the location
of the causal singularity. The causal singularity exists only when
$\beta^2_2 > 3(\nu^2-1)\alpha^2_2/4\nu^2$ and is hidden by the inner
horizon for all $\alpha_2$ and $\beta_2$ except when $|\alpha_2| =
|\beta_2|$ for which $r_-$ and $r_{s+}$ coincide. When $\beta^2_2
\leq 3(\nu^2-1)\alpha^2_2/4\nu^2$ one finds no CTCs upon
identification along $\eta_b$ and there are no singular regions of the spacetime.
\newline
\newline
Identifying along $\eta_c$ with $\beta_2>0$ gives us the extremal
black hole solutions of \ref{eq:ng} whereas with $\beta_2<0$ there
are CTCs close to the boundary. Identifying along $\eta_a$ with
$\alpha_0\neq0$ gives us naked causal singularities. Finally,
identifying along $\eta_a$ with $\alpha_0 = 0$ gives us no CTCs, and
we obtain the self-dual type solutions which we will now describe.

\subsection*{Self-Dual Solutions}
When we identify along $J_2$ we find no CTCs and one could call the
quotients \emph{self-dual} solutions in analogy to those quotients
of AdS$_3$ studied in \cite{Coussaert:1994tu}. In Schwarzschild
coordinates this corresponds to identifying $t$, i.e. $t \sim t +
2\pi\alpha$. Thus we can define a left moving temperature $T_L =
(\nu^2+3)\alpha/(4\pi\nu\ell)$ in the same spirit as \ref{tltr}. The
right moving temperature vanishes. The entropy of the Killing
horizon can be computed using the techniques in
\cite{Kraus:2005vz,Tachikawa:2006sz,Solodukhin:2005ah} and we
get:\footnote{In section \ref{sec:thermo} we will define the
quantity $c_L \equiv 4\nu\ell/(\nu^2+3)G$ in \ref{eq:ssd} as a left
moving central charge (together with a similar formula for $c_R$)
and see that the stretched black hole entropy of \ref{eq:ng} obeys
the Cardy formula. Using the same $c_L$ we see that the self-dual
solution has entropy obeying the Cardy formula as well. }%Furthermore,
%since the Hawking temperature as defined in \ref{eq:ht} vanishes for
%the self-dual solution, the first law of black hole thermodynamics
%is trivially satisfied, i.e. $d\mathcal{M}^{ADT} = -\Omega_H
%d\mathcal{J}^{ADT}$, where $\Omega_H$ is the angular velocity at the
%horizon defined in \ref{eq:hang}.

\begin{equation}
S = S_E + S_{CS} = \frac{\pi\alpha\ell}{3G} =
\frac{\pi^2T_L\ell}{3}\frac{4\nu\ell}{(\nu^2+3)G} \equiv \frac{\pi^2\ell}{3}c_LT_L \label{eq:ssd}
\end{equation}
%The conserved ADT charges of this solution \cite{Bouchareb:2007yx}
%with respect to the $\partial_\phi$ and $\partial_t$ Killing vectors
%are given by

%\begin{eqnarray}
%\mathcal{M}&=&{\ell\nu^2(\nu^2+3)\over18G(\nu^2-1)}(1-{\sqrt{\nu^2+3}\over2\nu})^2\alpha r_h^2\label{msd}\\
%\mathcal{J}&=&-{\ell\nu(\nu^2+3)\over9G(\nu^2-1)}(1-{\sqrt{\nu^2+3}\over2\nu})\alpha^2r_h\label{jsd}
%\end{eqnarray}

Notice that in Schwarzschild coordinates \ref{eq:ng}, the self-dual solution has a Killing horizon at the value of $r$  where $g^{rr}$ vanishes. Thus, even though these objects don't fall strictly under the category of regular black holes as defined above, they are in many ways like black holes including their thermodynamic behavior.
 A similar situation holds for all the self-dual black holes we uncover in what follows. These ones, along with their $\nu<1$ squashed counterparts, are however of particular interest as they appear as the near horizon region of extremal Kerr at fixed polar angle \cite{Bardeen:1999px, Bengtsson:2005zj}.

\subsubsection{Squashed}
In the \emph{squashed} case with $\nu<1$ the norms of the Killing vectors are negative at the boundary unless we identify along $\eta_a$ with $\alpha_0=0$ or along $\eta_b$ with $\alpha_2=0$. Then the quotients become the self-dual type solutions we mentioned.

The entropy of the squashed self-dual solutions is again given by
\ref{eq:ssd}.
% and conserved charges with respect to the $\partial_t$ and $\partial_\phi$ Killing vectors are given by \ref{msd} and \ref{jsd} respectively.
Notice that since we are identifying the $\partial_u$ direction,
which is always spacelike for spacelike warped AdS, there is no
pathology associated with the self-dual solution, even for $\nu<1$.
As we saw in \ref{eq:ssd}, such self-dual solutions have an entropy
obeying the Cardy formula.
%and conserved charges given by \ref{msd} and \ref{jsd} obeying the first law.
Furthermore, they have a Killing horizon in the Schwarzschild
coordinates. All these facts suggest that they should be regarded as
black holes.

\subsection{Timelike Warped}
The one-parameter subgroups are given by
\begin{eqnarray}
\xi_a:&& \alpha_0 \tilde{J}_0+\beta_2 J_2\\
\xi_b:&& \alpha_0\tilde{J}_0+\beta_0 J_0\\
\xi_c:&& \alpha_0 \tilde{J}_0+ J_0+J_2
\end{eqnarray}
The norms of the Killing vectors generating the subgroups are given in \ref{eq:tnorm1}, \ref{eq:tnorm2} and \ref{eq:tnorm3}.
\subsubsection{Stretched}
In the \emph{stretched} case we already encountered CTCs at infinity for the vacuum solution without any quotienting. Hence there are no regular black holes.
\subsubsection{Squashed}
For the \emph{squashed} case, identifying along $\xi_a$ we find no horizons. When $\alpha_0^2<3(1-\nu^2)\beta^2_2/4\nu^2$ we find no CTCs either. Identifying along $\xi_b$ always gives rise to CTCs outside the horizon. Identifying along $\xi_c$ always gives rise to CTCs outside the horizon unless $\alpha_0$ vanishes in which case we have no CTCs and we have a self-dual type solution.

\subsection{Null Warped}
The one-parameter subgroups are given by
\begin{eqnarray}
n_a&=&\alpha_0N_0+N\\
n_b&=&\alpha_1N_1+N\\
n_c&=&\alpha_1(N_1+N_0)+N
\end{eqnarray}
The norms of the Killing vectors generating the subgroups are given in \ref{eq:nnorm2}, \ref{eq:nnorm1} and \ref{eq:nnorm3}.
\subsubsection{Minus Sign}
For null warped AdS$_3$ with a \emph{minus} sign we encounter CTCs at the boundary unless $\alpha_0=0$ in which case we have a singularity-free solution.
\subsubsection{Plus Sign}
For null warped AdS$_3$ with a \emph{plus} sign, identifying along
$n_b$ gives no causal singularity when $\alpha_a$ is positive and
gives CTCs outside the horizon when $\alpha_1$ is negative.
\newline
\newline
Identifying along $n_a$ we encounter a horizon and causal singularity located at
\begin{equation}
r_h = 0, \quad r_s = \frac{-1\pm\sqrt{1-\alpha_0^2}}{2}
\end{equation}
The horizon, $r_h$, lies outside the causal singularity, $r_s$, whenever $0<\alpha_0<1$. One can obtain a real coordinate transformation taking the quotiented null warped metric to the following metric
\begin{equation}
{ds^2\over\ell^2}=-{r^2\over r^2+ r+{{\tilde{\alpha}_0}^2}}dt^2+\left(r^2+ r+{{\tilde{\alpha}_0}^2}\right)\left(d\phi-{r dt\over r^2+
r+{{\tilde{\alpha}_0}^2}}\right)^2+{dr^2\over4r^2}
\end{equation}
where $r\in[r_s,\infty]$, $u\in[-\infty,\infty]$ and $\phi \sim \phi
+ 2\pi$ and we have defined $\tilde{\alpha}_0 \equiv \alpha_0/2$.
The coordinate transformation is given by
\begin{eqnarray}
x^-&=&e^{\alpha_0\phi}\\
x^+&=& \phi-2t-{\alpha_0\over2 r}\\
u&=&\sqrt{{\alpha_0e^{\alpha_0\phi}\over  r}}
\end{eqnarray}
We can recognize the above solution as the extremal BC black hole
\ref{eq:bc} for the special value $\Omega=\frac{1}{2}$ and $\rho_0 =
3(\nu^2-1)\alpha_0^2/4(\nu^2+3)$ in the limit $\nu \to 1$. The
conserved ADT charges \cite{Bouchareb:2007yx} of this black hole
with respect to the $\partial_t$ and $\partial_\phi$ Killing vectors
are given by
\begin{eqnarray}
\mathcal{M}^{ADT}=0, \quad \mathcal{J}^{ADT}=-{\alpha_0^2\ell\over12G}
\end{eqnarray}
The temperature and angular momentum at the horizon as defined in
\ref{eq:ht} and \ref{eq:hang} vanish for the above solution. However
the right moving temperature is non-zero:
\begin{equation} T_R =
\frac{\alpha_0}{2\pi \ell}
\end{equation} and the thermodynamic entropy
is given by
\begin{equation}
S = \frac{\pi\alpha_0\ell}{3G} = \frac{\pi^2\ell
T_R}{3}\frac{(5\nu^2+3)\ell}{G\nu(\nu^2+3)}\mid_{\nu \to 1}
\end{equation}
The term $(5\nu^2+3)\ell/G\nu(\nu^2+3)$ will be
recognized as the right moving central charge in the following
section. A similar situation to $n_a$ holds if we identify along
$n_c$.

\section{Thermodynamics}
\label{sec:thermo}
The thermodynamics of spacelike stretched black holes was studied in \cite{Bouchareb:2007yx} where it was shown that, after accounting for the effects of the Chern-Simons term on the various thermodynamic quanitities,  they obey the first law. In this section we reorganize the formulae in a suggestive manner to motivate the conjecture of the concluding section.
\subsection{Entropy}
The entropy of the warped black hole is composed of two terms. There is the usual term stemming from the Einstein action and a term coming from the Chern-Simons part of the action \cite{Kraus:2005vz,Tachikawa:2006sz,Solodukhin:2005ah}. The total entropy of the warped black hole is given by
\begin{equation}
S={\pi\ell\over24\nu G}\left[(9\nu^2+3){r}_+-(\nu^2+3){r}_--4\nu\sqrt{(\nu^2+3){r}_+{r}_-}\right]
\end{equation}
It is instructive to rewrite this in terms of the temperatures $T_L$ and $T_R$ appearing in \ref{tltr}.
Defining right and left moving ``central charges",
\begin{eqnarray}
c_R &\equiv&   {(5\nu^2+3)\ell\over G\nu(\nu^2+3)}\label{cc2r}\\
c_L &\equiv&   {4\nu\ell\over G(\nu^2+3)}
\label{cc2l}
\end{eqnarray}
allows us to express the entropy of the warped black hole in the following suggestive manner
\begin{equation}
S = \frac{\pi^2\ell}{3}\left( c_L T_L + c_R T_R \right).
\end{equation}
This is precisely the formula for the entropy of a 2d
CFT with central charges $c_L$ and $c_R$ at temperatures $T_L$ and
$T_R$. Of course the central charges were defined so that this
relation holds: a nontrivial observation is that so defined they
depend only on the coupling constant $\nu$ and not ${r}_\pm$. A
further nontrivial fact is  that the left and right moving central
charges, $c_L$ and $c_R$, are equal to the left and right moving
central charges of TMG in AdS$_3$ when $\nu=1$. A final nontrivial
check is that the difference between the left and right moving
central charges matches the coefficient of the diffeomorphism
anomaly \cite{Kraus:2005zm} \be c_L - c_R = -\frac{\ell}{G\nu} \ee
One can also define the following left and right moving energies,
\begin{eqnarray}
E_L &\equiv& \frac{\pi^2\ell}{6}c_LT_L^2\\
E_R &\equiv& \frac{\pi^2\ell}{6}c_RT_R^2
\end{eqnarray}
Which by construction obey
\begin{eqnarray}
\frac{dS}{dE_L}&=& \frac{1}{T_L}\\
\frac{dS}{dE_R}&=& \frac{1}{T_R}.
\label{flw}
\end{eqnarray}

\subsection{First law}
A black hole is characterized by conserved charges such as the energy or angular momentum, as well as conjugate variables such as temperature or angular potential.  It is a nontrivial test of black hole dynamics that these quantities are related by the first law.  The spacelike stretched black holes were shown to pass this test in  \cite{Bouchareb:2007yx}. They compute the so-called ADT mass $\mathcal{M}^{ADT}$ and angular momentum $\mathcal{J}^{ADT}$, using the surface integral expressions derived in \cite{Abbott:1982jh,Deser:2002rt,Deser:2002jk,Deser:2003vh}
for the conserved charges associated to the asymptotic Killing vectors $\partial_t$ and $\partial_\theta$ of  \ref{eq:ng} and find
\begin{equation}
\mathcal{M}^{ADT} =  {(\nu^2+3)\over24 G}\left(\nuo\right)
\end{equation}
\begin{multline}
\mathcal{J}^{ADT} =  \frac{\nu\ell(\nu^2 + 3)}{96 G}\left[\left(\nuo\right)^2\right.\\
\left.- \frac{(5\nu^2+3)}{4\nu^2}({r}_+-{r}_-)^2\right]
\end{multline}
They further compute the Hawking temperature $T_H$, defined as the surface gravity of the horizon divided by $2\pi$, and the angular velocity of the horizon $\Omega_H$  as
\begin{eqnarray}
T_H &\equiv& {1\over2\pi\ell}\sqrt{g^{{r}{r}}}\p_{r} N= {(\nu^2+3)\over4\pi\ell}{({r}_+-{r}_-)\over(2\nu{r}_+-\sqrt{(\nu^2+3){r}_+{r}_-})}\label{eq:ht}\\
\Omega_H &\equiv& \frac{N^\theta}{\ell} = {2\over(2\nu{r}_+-\sqrt{(\nu^2+3){r}_+{r}_-})\ell}\label{eq:hang}
\end{eqnarray}
They then explicitly check that these are related to the entropy via the first law:
\begin{equation}
T_H = \left({\partial S\over\partial \mathcal{M}^{ADT}}\right)^{-1}, \quad \Omega_H = T_H \left(\frac{\partial S}{\partial \mathcal{J}^{ADT}}\right).\label{flaw}
\end{equation}

In the preceding subsection we propose that the conserved charges $(E_L,E_R)$ and potentials $(T_L,T_R)$ are more natural for this system than
$(\mathcal{M}^{ADT},\mathcal{J}^{ADT})$ and $T_H, \Omega_H$. Since this
is just a change of variables - albeit an illuminating one - the first law will still hold for the new charges/potentials. To see this explicitly
one can express the conserved ADT charges in terms of the left and right moving temperatures and energies as follows,
\begin{eqnarray}
\mathcal{M}^{ADT} &=&{1 \over G}\sqrt{{2\ell E_L\over 3c_L}}\\
\mathcal{J}^{ADT} &=& \ell (E_L - E_R).
\end{eqnarray}
The potentials are related by \begin{eqnarray}
{1\over T_H}&=&{4\pi\nu\ell\over\nu^2+3}{T_L+T_R\over T_R}
\\
{\Omega_H\over T_H}
&=&{1\over T_R\ell}.\end{eqnarray}
One may then readily use these relations to  verify that \ref{flaw} is equivalent to the first law \ref{flw} in the new variables.\footnote{We note that similar formulae work for the null black holes which arise in the $\nu\to 1$ limit and have $T_L$=0.}

\subsection{A conjecture}
Topologically massive gravity  with $\nu >1$ ($\mu\ell>3$) admits
an
$U(1)_L\times SL(2,\mathbb{R})_R$-invariant candidiate ground state referred to as spacelike stretched (warped)  AdS$_3$, as well as regular black hole
solutions. We conjecture that with suitable asymptotically
stretched  AdS$_3$ boundary conditions $\nu>1$ quantum TMG
is holographically dual to
a 2D boundary CFT with $c_R =  {(5\nu^2+3)\ell\over G\nu(\nu^2+3)}$
and $c_L ={4\nu\ell\over G(\nu^2+3)}$.

The primary motivations for this conjecture are that application of
the Cardy formula to the CFT density of states reproduces the black
hole entropy, and that as far as we understand the  conjecture does
not contradict any known properties of the theory. The conjecture passed some weak tests in section 7. Further tests of
the conjecture are possible. After understanding the boundary
conditions, perturbative stability must be demonstrated. These
boundary conditions will also determine the asymptotic symmetry
group, and perhaps enable a check of the formulae \ref{cc2r} and
\ref{cc2l} for the central charges \cite{thas} along the lines of \cite{Compere:2007in}. Finally, it might be possible to engineer  TMG on
warped AdS$_3$ in a decoupling limit of string theory and find the dual CFT.

\section*{Acknowledgements}
This work was partially funded by an DOE grant DE-FG02-91ER40654. We
wish to thank M. Guica and T. Hartman for useful comments on the
draft. W. S. thanks the High Energy Group at Harvard for their kind
hospitality. M. P. is supported by an NSF fellowship.

\appendix
\setcounter{equation}{0}  % reset counter
\section{Isometries}  % use *-form to suppress numbering
\label{section:appA}

We write down the relevant Killing vectors for the various warped
AdS$_3$ metrics. The $SL(2,\mathbb{R})_L$ isometries are given by
\begin{eqnarray}
J_1 &=& {\frac{2 \sinh{u}}{\cosh{\sigma}} \partial_{\tau}-2\cosh{u} \partial_{\sigma}+2\tanh{\sigma} \sinh{u} \partial_{u}} \\
J_2 &=& 2\partial_{u} \\
J_0 &=& {-\frac{2 \cosh{u}}{\cosh{\sigma}} \partial_{\tau}+2\sinh{u} \partial_{\sigma}-2\tanh{\sigma} \cosh{u} \partial_{u}}
\end{eqnarray}
These satisfy the algebra $[J_1,J_2]=2J_0$, $[J_0,J_1]=-2J_2$ and $[J_0,J_2]=2J_1$. The $SL(2,\mathbb{R})_R$ isometries are given by
\begin{eqnarray}
\tilde{J}_1 &=&  {2\sin{\tau} \tanh{\sigma} \partial_{\tau} - 2\cos{\tau} \partial_{\sigma}+\frac{2\sin{\tau}}{\cosh{\sigma}} \partial_{u} }  \\
\tilde{J}_2 &=&  {-2\cos{\tau} \tanh{\sigma} \partial_{\tau}-2\sin{\tau} \partial_{\sigma} - \frac{2\cos{\tau}}{\cosh{\sigma}} \partial_{u}  }  \\
\tilde{J}_0 &=&  2\partial_{\tau}
\end{eqnarray}
These satisfy the algebra $[\tilde{J}_1,\tilde{J}_2]=2\tilde{J}_0$, $[\tilde{J}_0,\tilde{J}_1]=-2\tilde{J}_2$ and $[\tilde{J}_0,\tilde{J}_2]=2\tilde{J}_1$.
\newline
\newline
As mentioned in the text the Killing vectors preserved by spacelike
warped anti-de Sitter space are given by the $SL(2,\mathbb{R})_R$
and $J_2$. The Killing vectors preserved by timelike warped anti-de
Sitter space are given by the $SL(2,\mathbb{R})_L$ and
$\tilde{J}_0$.
\newline
\newline
The Killing vectors for null warped anti-de Sitter space are given by
\begin{eqnarray}
N_1&=&\p_-,\\
N_0&=&x^-\p_-+{u\over2}\p_u,
\\N_{-1}&=&(x^-)^2\p_--u^2\p_++x^-u\p_u
\\ N&=&\p_+
\end{eqnarray}
where the $U(1)_{null}$ is spanned by $N$. These satisfy the algebra $[N_1,N_0] = N_{1}$, $[N_{-1},N_0]=-N_{-1}$ and $[N_1,N_{-1}]= 2N_0$.

\subsection{Classification of $SL(2,\mathbb{R})\times U(1)$ Subgroups}

In this section we classify the various one-parameter subgroups of
the isometries for the three types of warped anti-de Sitter space.
Our classification is based on the classification of the
one-parameter subgroups of $SL(2,\mathbb{R})_L\times
SL(2,\mathbb{R})_R$ given in \cite{Banados:1992gq}.

\subsubsection{Spacelike Warped anti-de Sitter Space}

Spacelike warped AdS has isometry group $U(1)_L\times SL(2,\mathbb{R})_R$. The most general form of the Killing vector along which we
can quotient is
\begin{equation}
\beta_2 J_2+ \alpha_2 \tilde{J}_2+\alpha_0 \tilde{J}_0+\alpha_1\tilde{J}_1 .
\end{equation}
We can always get rid of the $\alpha_1$ via a
$U(1)_L\times SL(2,\mathbb{R})_R$ transformation. There are three
types of one parameter subgroups generated by
\begin{eqnarray}
\eta_a:&& \beta_2 J_2+\alpha_0 \tilde{J}_0\\
\eta_b:&& \beta_2 J_2+\alpha_2 \tilde{J}_2\\ \eta_c:&& \beta_2 J_2+ \tilde{J}_0+\tilde{J}_2
\end{eqnarray}
In terms of the classification in \cite{Banados:1992gq}, these
correspond to $I_a,~ I_b,$ and $II_a$ (the first form) respectively.
The norms of the above generators are
\begin{eqnarray}
\nonumber\eta_a^2&=&{12\ell^2(\nu^2-1)\over(\nu^2+3)^2}\left[\alpha_0\sinh(\sigma)\right]^2+{32\nu^2\ell^2\over(\nu^2+3)^2}\beta_2\left[\alpha_0\sinh(\sigma)\right]\\
&-&{4\ell^2\over\nu^2+3}\alpha_0^2+{16\nu^2\ell^2\over(\nu^2+3)^2}\beta_2^2\label{eq:snorm1}\\
\nonumber\eta_b^2&=&{12\ell^2(\nu^2-1)\over(\nu^2+3)^2}\left[\alpha_2\cos(\tau)\cosh(\sigma)\right]^2
-{32\nu^2\ell^2\over(\nu^2+3)^2}\beta_2\left[\alpha_2\cos(\tau)\cosh(\sigma)\right]\\
&+&{4\ell^2\over\nu^2+3}(\alpha_2^2)+{16\nu^2\ell^2\over(\nu^2+3)^2}\beta_2^2\label{eq:snorm2}\\
\nonumber\eta_c^2&=&{12\ell^2(\nu^2-1)\over(\nu^2+3)^2}\left[\sinh(\sigma)-\cos(\tau)\cosh(\sigma)\right]^2+{32\nu^2\ell^2\over(\nu^2+3)^2}\beta_2\left[\sinh(\sigma)-\cos(\tau)\cosh(\sigma)\right]\\
&+&{16\nu^2\ell^2\over(\nu^2+3)^2}\beta_2^2\label{eq:snorm3}
\end{eqnarray}

\subsubsection{Timelike Warped anti-de Sitter Space}
For timelike warped, the most general form of the Killing vector is
\begin{equation}
\alpha_0 \tilde{J}_0+\beta_2 J_2+\beta_0 J_0+\beta_1J_1 .
\end{equation}
We can always eliminate $\beta_1$ via an $SL(2,\mathbb{R})_L \times U(1)_R$ transformation. There are three types of one parameter subgroups generated by
\begin{eqnarray}
\xi_a:&& \alpha_0 \tilde{J}_0+\beta_2 J_2\\
\xi_b:&& \alpha_0 \tilde{J}_0+\beta_0 J_0\\
\xi_c:&& \alpha_0 \tilde{J}_0+ J_0+J_2
\end{eqnarray}
In terms of the classification in \cite{Banados:1992gq}, these correspond to $I_a,~ I_b,$ and $II_a$ (the first form) respectively. The norms are
\begin{eqnarray}
\nonumber \xi_a^2&=&-{12\ell^2(\nu^2-1)\over(\nu^2+3)^2}\left[\beta_2\sinh(\sigma)\right]^2
-{32\nu^2\ell^2\over(\nu^2+3)^2}\alpha_0\left[\beta_2\sinh(\sigma)\right]\\
&+&{4\ell^2\over\nu^2+3}(\beta_2^2)-{16\nu^2\ell^2\over(\nu^2+3)^2}\alpha_0^2\label{eq:tnorm1}\\
\nonumber \xi_b^2&=&-{12\ell^2(\nu^2-1)\over(\nu^2+3)^2}\left[\beta_0\cosh(u)\cosh(\sigma)\right]^2
+{32\nu^2\ell^2\over(\nu^2+3)^2}\alpha_0\left[\beta_0\cosh(u)\cosh(\sigma)\right]\\
&-&{4\ell^2\over\nu^2+3}(\beta_0^2)-{16\nu^2\ell^2\over(\nu^2+3)^2}\alpha_0^2\label{eq:tnorm2}\\
\nonumber \xi_c^2&=&-{12\ell^2(\nu^2-1)\over(\nu^2+3)^2}\left[\sinh(\sigma)-\cosh(u)\cosh(\sigma)\right]^2-{32\nu^2\ell^2\over(\nu^2+3)^2}\alpha_0\left[\sinh(\sigma)-\cosh(u)\cosh(\sigma)\right]\\
&-&{16\nu^2\ell^2\over(\nu^2+3)^2}\alpha_0^2\label{eq:tnorm3}
\end{eqnarray}

\subsubsection{Null Warped anti-de Sitter Space}

The most general linear combination of Killing vectors is
\begin{equation}
\xi= N+ \alpha_1 N_1+\alpha_0 N_0+\alpha_{-1}N_{-1}
\end{equation}
By a shift of $x^-$, we can eliminate the term with $N_{-1}$. Then
the distinct subgroups are generated by
\begin{eqnarray}
n_a&=&\alpha_0N_0+N\\
n_b&=&\alpha_1N_1+N\\
n_c&=&\alpha_1(N_1+N_0)+N
\end{eqnarray}
We have not included a coefficient in front of the $N$ because it is null. The norm is
\begin{eqnarray}
n_a^2&=&\ell^2\left(\pm {\alpha_0^2(x^-)^2\over u^4}+{\alpha_0\over u^2}(x^-)+{\alpha_0^2\over4} \right)\label{eq:nnorm2}\\
n_b^2&=&\ell^2\left(\pm {\alpha_1^2\over u^4}+{\alpha_1\over u^2}\right)\label{eq:nnorm1}\\
n_c^2&=&\ell^2\left(\pm{\alpha_1^2\over u^4}(1+x^-)^2+{\alpha_1\over u^2}(1+ x^-)+{\alpha_1^2\over4}\right)\label{eq:nnorm3}
\end{eqnarray}
where the plus sign corresponds to the null warped metric with a plus sign and vice versa.

\section{Extremal Black Holes}
\label{section:appB}

It is convenient to introduce spacelike warped AdS$_3$ in Poincare coordinates
\begin{equation}
{ds^2\over\ell^2}={1\over\nu^2+3}\left(-x^2d\psi^2+{dx^2\over
x^2}+{4\nu^2\over\nu^2+3}(d\phi+xd\psi)^2\right)
\end{equation}
and note that the coordinate transformation from Poincare
coordinates to the global coordinate in \ref{eq:ssqads} is:
\begin{eqnarray}
\psi&=&-{\sin(\tau)\over2(\tanh(\sigma)-\cos(\tau))}\\
\phi&=&u+2\tanh^{-1}\left(e^\sigma\tan({\tau\over2})\right)\\
x&=&2(\sinh(\sigma)-\cos(\tau)\cosh(\sigma))
\end{eqnarray}
The metric of extremal warped black hole is obtained by setting
$r_+=r_-=r_h$ in \ref{eq:ng} to obtain
\begin{eqnarray}
\frac{ds^2}{\ell^2}&=&dt^2+\frac{ d{r}^2}{(\nu^2+3)({r}-{r}_{h})^2} + \left(2\nu {r}  - \sqrt{\nu^2+3}r_h\right)dtd\theta \\
\nonumber\\&&+\frac{{r}}{4}\left(3(\nu^2 - 1){r}+2(\nu^2+3){r}_h -
4\nu r_h\sqrt{\nu^2+3}\right)d\theta^2 \label{eq:ebh}
\end{eqnarray}
Then the coordinate transformation from the Poincare coordinate to the extremal black hole is
\begin{eqnarray}
t&=&{2\nu\over(\nu^2+3)}\phi+{r_h(-2\nu+\sqrt{\nu^2+3})\psi\over(\nu^2+3)} \\
\theta&=&{2\psi\over\nu^2+3}\\
r&=&x+r_h
\end{eqnarray}
Combining the above coordinate transformations gives the coordinate
transformation between warped AdS$_3$ and the extremal warped black
hole,
\begin{eqnarray}
t&=&{2\nu\over(\nu^2+3)}\left(u+2\tanh^{-1}\left(e^\sigma\tan({\tau\over2})\right)\right)
-{r_h(-2\nu+\sqrt{\nu^2+3})\sin(\tau)\over2(\nu^2+3)(\tanh(\sigma)-\cos(\tau))}\\
\theta&=&-{\sin(\tau)\over(\nu^2+3)(\tanh(\sigma)-\cos(\tau))}\\
r&=&2(\sinh(\sigma)-\cos(\tau)\cosh(\sigma))+r_h
\end{eqnarray}

\section{BC Coordinates}
\label{sec:bc}
\subsection{Metric}

In \cite{Bouchareb:2007yx} a warped black hole closely related to the Schwarzschild solution in \ref{eq:ng} was obtained by a simple transformation which we describe below. We will refer to these coordinates as BC coordinates for which the metric is given by
\begin{equation}
\frac{ds^2_{BC}}{\ell^2} = - \frac{\left(\nu^2+3\right)}{4\nu^2} \frac{\rho^2-\rho^2_0}{R(\rho)^2}dT^2+\frac{1}{\left(\nu^2+3\right)}\frac{d\rho^2}{\rho^2-\rho^2_0}+R(\rho)^2\left[
d\phi - \frac{4\nu^2\rho+3(\nu^2-1)\Omega}{4\nu^2R(\rho)^2}dT \right]^2
\label{eq:bc}
\end{equation}
where $\rho\in [\rho_s,\infty]$, $T \in[-\infty,\infty]$ and $\phi \sim \phi + 2\pi$. We have defined
\begin{eqnarray}
R(\rho)^2 &\equiv& \rho^2 + 2\Omega\rho +
\frac{3\Omega^2(\nu^2-1)}{4\nu^2}+\frac{(\nu^2+3)}{3(\nu^2-1)}\rho^2_0
\end{eqnarray}
such that $R(\rho_s) = 0$. There is a vacuum solution analogous to the massless BTZ or Poincare anti-de Sitter space for the BC black hole which is given by setting $\rho_0=\Omega=0$. The horizons of the BC black hole are located at $\rho = \pm\rho_0$ where $\rho_0 > 0$ and the (causal) singularities are located at $\rho = \rho_s$. We would like to emphasize that these are singularities in the causal structure \cite{Banados:1992gq} and not spacetime singularities that appear in the coordinate invariants. The metric \ref{eq:bc} contains such causal singularities in the parameter range
\begin{equation}
\Omega^2 > \frac{4\nu^2 \rho^2_0}{3(\nu^2-1)} \label{eq:sing}
\end{equation}
In order to avoid naked singularities we require that
$\rho_0 > \rho_s$ which gives us the bound
\begin{equation}
\frac{3(\nu^2-1)}{4\nu^2}\left[\Omega+\frac{4\nu^2
\rho_0}{3(\nu^2-1)}\right]^2>0
\end{equation} This condition is
satisfied for $\nu^2 > 1$ and thus the physical black holes for both
the Schwarzschild and BC coordinates are in the same parameter
region. The location of the horizon and singularity coincide
whenever $\Omega = - 4\nu^2 \rho_0/3(\nu^2-1)$.

\subsection{Schwarzschild to BC Coordinate Transformation}
\label{sec:ngbc}

The coordinate transformation relating the solution in Schwarzschild coordinates \ref{eq:ng} to the BC solution \ref{eq:bc} is given by
\begin{eqnarray}
t&=& {1\over\nu}\left({3(\nu^2-1)\over 4}\right)^{1/2} T\\
\theta &=& -\left({4\over3(\nu^2-1)}\right)^{1/2}\phi\\
r &=& \rho + \frac{1}{2} (r_+ + r_-)
\label{eq:bcng}
\end{eqnarray}
%\begin{eqnarray}
%t&=& {3\over\mu}\left(3(\mu^2-9)\over (\mu^2+27)(r_+^2+r_-^2)-4\mu \sqrt{\mu^2+27} r_+ r_-\right)^{1/2} T\\
%\theta &=& -\left((\mu^2+27)(r_+^2+r_-^2)-4\mu \sqrt{\mu^2+27} r_+ r_-\over
%3(\mu^2-9)\right)^{1/2}\phi \label{eq:phi}\\
%r^2 &=& \rho  + \frac{1}{2} (r_+^2+r_-^2)
%\end{eqnarray}
We note that the transformation is singular when $\nu^2 = 1$. When $\nu^2 < 1$ the transformation becomes imaginary. This corresponds to the parameter space containing unphysical black holes. The relation between the parameters of the Schwarzschild solution and the parameters of the BC solution can also be obtained quite simply and is given by
\begin{eqnarray}
\rho_0^2&=&\frac{1}{4} (r_+ - r_-)^2\\
\Omega&=&\frac{2 \nu^2}{3(\nu^2-1)} \left(r_+ + r_- - \frac{\sqrt{r_+ r_-(\nu^2+3)}}{\nu} \right)
\end{eqnarray}
such that $\rho_s = -(r_+ + r_-)/2$.

%%%%%%%%%%%%%%%%%%%%%%%%%%%%%%%%%%%%%%%%%%%%%%%%%%
%%%%%%%%%%%%%%%%%%%%%%%%%%%%%%%%%%%%%%%%%%%%%%%%%%


\begin{thebibliography}{1}
\addcontentsline{toc}{section}{References}

%\cite{Deser:1981wh}
\bibitem{Deser:1981wh}
  S.~Deser, R.~Jackiw and S.~Templeton,
  ``Topologically massive gauge theories,''
  Annals Phys.\  {\bf 140}, 372 (1982)
  [Erratum-ibid.\  {\bf 185}, 406.1988\ APNYA,281,409 (1988\ APNYA,281,409-449.2000)].
  %%CITATION = APNYA,281,409;%%

  %\cite{Deser:1982vy}
\bibitem{Deser:1982vy}
  S.~Deser, R.~Jackiw and S.~Templeton,
  ``Three-Dimensional Massive Gauge Theories,''
  Phys.\ Rev.\ Lett.\  {\bf 48}, 975 (1982).
  %%CITATION = PRLTA,48,975;%%

\bibitem{Carlip:2008jk}
  S.~Carlip, S.~Deser, A.~Waldron and D.~K.~Wise,
  ``Cosmological Topologically Massive Gravitons and Photons,''
  arXiv:0803.3998 [hep-th].
  %%CITATION = ARXIV:0803.3998;%%

%\cite{Li:2008yz}
\bibitem{Li:2008yz}
  W.~Li, W.~Song and A.~Strominger,
  ``Comment on 'Cosmological Topological Massive Gravitons and Photons',''
  arXiv:0805.3101 [hep-th].
  %%CITATION = ARXIV:0805.3101;%%

%\cite{Brown:1986nw}
\bibitem{Brown:1986nw}
  J.~D.~Brown and M.~Henneaux,
  ``Central Charges in the Canonical Realization of Asymptotic Symmetries: An
  Example from Three-Dimensional Gravity,''
  Commun.\ Math.\ Phys.\  {\bf 104}, 207 (1986).

  %%CITATION = CMPHA,104,207;%%
\bibitem{fg}
C. Fefferman and C. R. Graham, "Conformal Invariants" {\it Elie
Cartan et les Mathematiques d'Aujordhui} (Asterique, 1985) 95.

  %\cite{Hotta:2008yq}
\bibitem{Hotta:2008yq}
  K.~Hotta, Y.~Hyakutake, T.~Kubota and H.~Tanida,
  ``Brown-Henneaux's Canonical Approach to Topologically Massive Gravity,''
  arXiv:0805.2005 [hep-th].
  %%CITATION = ARXIV:0805.2005;%%

 \bibitem{kw}
  I.~R.~Klebanov and E.~Witten,
  ``AdS/CFT correspondence and symmetry breaking,''
  Nucl.\ Phys.\  B {\bf 556}, 89 (1999)
  [arXiv:hep-th/9905104].
  %%CITATION = NUPHA,B556,89;%%


  %\cite{Grumiller:2008qz}
\bibitem{Grumiller:2008qz}
  D.~Grumiller and N.~Johansson,
  ``Instability in cosmological topologically massive gravity at the chiral
  point,''
  arXiv:0805.2610 [hep-th].
  %%CITATION = ARXIV:0805.2610;%%

%\cite{Li:2008dq}
\bibitem{Li:2008dq}
  W.~Li, W.~Song and A.~Strominger,
  ``Chiral Gravity in Three Dimensions,''
  JHEP {\bf 0804}, 082 (2008)
  [arXiv:0801.4566 [hep-th]].
  %%CITATION = JHEPA,0804,082;%%

%\cite{Vuorio:1985ta}
\bibitem{Vuorio:1985ta}
  I.~Vuorio,
  ``Topologically Massive Planar Universe,''
  Phys.\ Lett.\  B {\bf 163}, 91 (1985).
  %%CITATION = PHLTA,B163,91;%%

  %\cite{Percacci:1986ja}
\bibitem{Percacci:1986ja}
  R.~Percacci, P.~Sodano and I.~Vuorio,
 ``Topologically Massive Planar Universes With Constant Twist,''
  Annals Phys.\  {\bf 176}, 344 (1987).
  %%CITATION = APNYA,176,344;%%

  %\cite{Nutku:1993eb}
\bibitem{Nutku:1993eb}
  Y.~Nutku,
  ``Exact solutions of topologically massive gravity with a cosmological
  constant,''
  Class.\ Quant.\ Grav.\  {\bf 10}, 2657 (1993).
  %%CITATION = CQGRD,10,2657;%%

    %\cite{Gurses}
\bibitem{Gurses}
  M.~G\"urses,
  ``Perfect fluid sources in 2+1 dimensions,"
  Class.\ Quant.\ Grav.\  {\bf 11}, 2585 (1994).

  %\cite{Bouchareb:2007yx}
\bibitem{Bouchareb:2007yx}
  A.~Bouchareb and G.~Clement,
  ``Black hole mass and angular momentum in topologically massive gravity,''
  Class.\ Quant.\ Grav.\  {\bf 24}, 5581 (2007)
  [arXiv:0706.0263 [gr-qc]].
  %%CITATION = CQGRD,24,5581;%%

%\cite{Banados:1992wn}
\bibitem{Banados:1992wn}
  M.~Banados, C.~Teitelboim and J.~Zanelli,
  ``The Black hole in three-dimensional space-time,''
  Phys.\ Rev.\ Lett.\  {\bf 69}, 1849 (1992)
  [arXiv:hep-th/9204099].
  %%CITATION = PRLTA,69,1849;%%

  %\cite{Banados:1992gq}
\bibitem{Banados:1992gq}
  M.~Banados, M.~Henneaux, C.~Teitelboim and J.~Zanelli,
  ``Geometry of the (2+1) black hole,''
  Phys.\ Rev.\  D {\bf 48}, 1506 (1993)
  [arXiv:gr-qc/9302012].
  %%CITATION = PHRVA,D48,1506;%%

 %\cite{Coussaert:1994tu}
\bibitem{Coussaert:1994tu}
  O.~Coussaert and M.~Henneaux,
  ``Self-dual solutions of 2+1 Einstein gravity with a negative  cosmological
  constant,''
  arXiv:hep-th/9407181.
  %%CITATION = HEP-TH/9407181;%%



%\cite{Rooman:1  998xf}
\bibitem{Rooman:1998xf}
  M.~Rooman and P.~Spindel,
  ``Goedel metric as a squashed anti-de Sitter geometry,''
  Class.\ Quant.\ Grav.\  {\bf 15}, 3241 (1998)
  [arXiv:gr-qc/9804027].
  %%CITATION = CQGRD,15,3241;%%

%\cite{Duff:1998cr}
\bibitem{Duff:1998cr}
  M.~J.~Duff, H.~Lu and C.~N.~Pope,
  ``AdS(3) x S**3 (un)twisted and squashed, and an O(2,2,Z) multiplet of
  dyonic strings,''
  Nucl.\ Phys.\  B {\bf 544}, 145 (1999)
  [arXiv:hep-th/9807173].
  %%CITATION = NUPHA,B544,145;%%

%\cite{Israel:2003ry}
\bibitem{Israel:2003ry}
  D.~Israel, C.~Kounnas and M.~P.~Petropoulos,
  ``Superstrings on NS5 backgrounds, deformed AdS(3) and holography,''
  JHEP {\bf 0310}, 028 (2003)
  [arXiv:hep-th/0306053].
  %%CITATION = JHEPA,0310,028;%%

%\cite{Andrade:2005ur}
\bibitem{Andrade:2005ur}
  T.~Andrade, M.~Banados, R.~Benguria and A.~Gomberoff,
  ``The 2+1 charged black hole in topologically massive electrodynamics,''
  Phys.\ Rev.\ Lett.\  {\bf 95}, 021102 (2005)
  [arXiv:hep-th/0503095].
  %%CITATION = PRLTA,95,021102;%%

\bibitem{Bengtsson:2005zj}
  I.~Bengtsson and P.~Sandin,
  ``Anti-de Sitter space, squashed and stretched,''
  Class.\ Quant.\ Grav.\  {\bf 23}, 971 (2006)
  [arXiv:gr-qc/0509076].
  %%CITATION = CQGRD,23,971;%%

%\cite{Banados:2005da}
\bibitem{Banados:2005da}
  M.~Banados, G.~Barnich, G.~Compere and A.~Gomberoff,
  ``Three dimensional origin of Goedel spacetimes and black holes,''
  Phys.\ Rev.\  D {\bf 73}, 044006 (2006)
  [arXiv:hep-th/0512105].
  %%CITATION = PHRVA,D73,044006;%%

 %\cite{Son:2008ye}
\bibitem{Son:2008ye}
  D.~T.~Son,
  ``Toward an AdS/cold atoms correspondence: a geometric realization of the
  Schroedinger symmetry,''
  arXiv:0804.3972 [hep-th].
  %%CITATION = ARXIV:0804.3972;%%

  %\cite{Balasubramanian:2008dm}
\bibitem{Balasubramanian:2008dm}
  K.~Balasubramanian and J.~McGreevy,
  ``Gravity duals for non-relativistic CFTs,''
  arXiv:0804.4053 [hep-th].
  %%CITATION = ARXIV:0804.4053;%%


\bibitem{Bardeen:1999px}
  J.~M.~Bardeen and G.~T.~Horowitz,
  ``The extreme Kerr throat geometry: A vacuum analog of AdS(2) x S(2),''
  Phys.\ Rev.\  D {\bf 60}, 104030 (1999)
  [arXiv:hep-th/9905099].
  %%CITATION = PHRVA,D60,104030;%%

\bibitem{ghs}
M. Guica, T. Hartman and A. Strominger, in progress.


%\cite{Strominger:1997eq}
\bibitem{Strominger:1997eq}
  A.~Strominger,
  ``Black hole entropy from near-horizon microstates,''
  JHEP {\bf 9802}, 009 (1998)
  [arXiv:hep-th/9712251].
  %%CITATION = JHEPA,9802,009;%%

 \bibitem{cdww}
  S.~Carlip, S.~Deser, A.~Waldron and D.~K.~Wise,
  ``Topologically Massive AdS Gravity,''
  arXiv:0807.0486 [hep-th].

  \bibitem{gps}
  G.~W.~Gibbons, C.~N.~Pope and E.~Sezgin,
  ``The General Supersymmetric Solution of Topologically Massive
  Supergravity,''
  arXiv:0807.2613 [hep-th].

\bibitem{Adams:2008wt}
  A.~Adams, K.~Balasubramanian and J.~McGreevy,
  ``Hot Spacetimes for Cold Atoms,''
  arXiv:0807.1111 [hep-th].

 %\cite{Herzog:2008wg}
\bibitem{Herzog:2008wg}
  C.~P.~Herzog, M.~Rangamani and S.~F.~Ross,
  ``Heating up Galilean holography,''
  arXiv:0807.1099 [hep-th].
  %%CITATION = ARXIV:0807.1099;%%

%\cite{Maldacena:2008wh}
\bibitem{Maldacena:2008wh}
  J.~Maldacena, D.~Martelli and Y.~Tachikawa,
  ``Comments on string theory backgrounds with non-relativistic conformal
  %symmetry,''
  arXiv:0807.1100 [hep-th].
  %%CITATION = ARXIV:0807.1100;%%


%\cite{Israel:2004vv}
\bibitem{Israel:2004vv}
  D.~Israel, C.~Kounnas, D.~Orlando and P.~M.~Petropoulos,
  ``Electric / magnetic deformations of S**3 and AdS(3), and geometric
  cosets,''
  Fortsch.\ Phys.\  {\bf 53}, 73 (2005)
  [arXiv:hep-th/0405213].
  %%CITATION = FPYKA,53,73;%%

    %\cite{Bengtsson:2005zj}


   %\cite{Detournay:2005fz}
\bibitem{Detournay:2005fz}
  S.~Detournay, D.~Orlando, P.~M.~Petropoulos and P.~Spindel,
  ``Three-dimensional black holes from deformed anti de Sitter,''
  JHEP {\bf 0507}, 072 (2005)
  [arXiv:hep-th/0504231].
  %%CITATION = JHEPA,0507,072;%%


    %\cite{Compere:2007in}
\bibitem{Compere:2007in}
  G.~Compere and S.~Detournay,
  ``Centrally extended symmetry algebra of asymptotically Goedel spacetimes,''
  JHEP {\bf 0703}, 098 (2007)
  [arXiv:hep-th/0701039].
  %%CITATION = JHEPA,0703,098;%%

 %\cite{Strominger:1998yg}
\bibitem{Strominger:1998yg}
  A.~Strominger,
  ``AdS(2) quantum gravity and string theory,''
  JHEP {\bf 9901}, 007 (1999)
  [arXiv:hep-th/9809027].
  %%CITATION = JHEPA,9901,007;%%
    %\cite{Moussa:2003fc}
\bibitem{Moussa:2003fc}
  K.~A.~Moussa, G.~Clement and C.~Leygnac,
  ``The black holes of topologically massive gravity,''
  Class.\ Quant.\ Grav.\  {\bf 20}, L277 (2003)
  [arXiv:gr-qc/0303042].
  %%CITATION = CQGRD,20,L277;%%

    %\cite{Abbott:1982jh}
\bibitem{Abbott:1982jh}
  L.~F.~Abbott and S.~Deser,
  ``Charge Definition In Nonabelian Gauge Theories,''
  Phys.\ Lett.\  B {\bf 116}, 259 (1982).
  %%CITATION = PHLTA,B116,259;%%

  %\cite{Deser:2002rt}
\bibitem{Deser:2002rt}
  S.~Deser and B.~Tekin,
  ``Gravitational energy in quadratic curvature gravities,''
  Phys.\ Rev.\ Lett.\  {\bf 89}, 101101 (2002)
  [arXiv:hep-th/0205318].
  %%CITATION = PRLTA,89,101101;%%

  %\cite{Deser:2002jk}
\bibitem{Deser:2002jk}
  S.~Deser and B.~Tekin,
  ``Energy in generic higher curvature gravity theories,''
  Phys.\ Rev.\  D {\bf 67}, 084009 (2003)
  [arXiv:hep-th/0212292].
  %%CITATION = PHRVA,D67,084009;%%

  %\cite{Deser:2003vh}
\bibitem{Deser:2003vh}
  S.~Deser and B.~Tekin,
  ``Energy in topologically massive gravity,''
  Class.\ Quant.\ Grav.\  {\bf 20}, L259 (2003)
  [arXiv:gr-qc/0307073].
  %%CITATION = CQGRD,20,L259;%%

  \bibitem{Sousa}
  F.~C.~Sousa, J.~B.~Fonseca, and C.~Romero,
  ``Equivalence of three-dimensional spacetimes,"
  Class.\ Quant. \ Grav. \  {\bf 25}, 035007 (2008)

\bibitem{Maldacena:1998bw}
  J.~M.~Maldacena and A.~Strominger,
  ``AdS(3) black holes and a stringy exclusion principle,''
  JHEP {\bf 9812}, 005 (1998)
  [arXiv:hep-th/9804085].
  %%CITATION = JHEPA,9812,005;%%

  %\cite{Spradlin:1999bn}
\bibitem{Spradlin:1999bn}
  M.~Spradlin and A.~Strominger,
  ``Vacuum states for AdS(2) black holes,''
  JHEP {\bf 9911}, 021 (1999)
  [arXiv:hep-th/9904143].
  %%CITATION = JHEPA,9911,021;%%

\bibitem{Kraus:2005vz}
  P.~Kraus and F.~Larsen,
  ``Microscopic black hole entropy in theories with higher derivatives,''
  JHEP {\bf 0509}, 034 (2005)
  [arXiv:hep-th/0506176].
  %%CITATION = JHEPA,0509,034;%%

%\cite{Tachikawa:2006sz}
\bibitem{Tachikawa:2006sz}
  Y.~Tachikawa,
  ``Black hole entropy in the presence of Chern-Simons terms,''
  Class.\ Quant.\ Grav.\  {\bf 24}, 737 (2007)
  [arXiv:hep-th/0611141].
  %%CITATION = CQGRD,24,737;%%

  %\cite{Solodukhin:2005ah}
\bibitem{Solodukhin:2005ah}
  S.~N.~Solodukhin,
  ``Holography with gravitational Chern-Simons,''
  Phys.\ Rev.\  D {\bf 74}, 024015 (2006)
  [arXiv:hep-th/0509148].
  %%CITATION = PHRVA,D74,024015;%%

%\cite{Kraus:2005zm}
\bibitem{Kraus:2005zm}
  P.~Kraus and F.~Larsen,
  ``Holographic gravitational anomalies,''
  JHEP {\bf 0601}, 022 (2006)
  [arXiv:hep-th/0508218].
  %%CITATION = JHEPA,0601,022;%%

\bibitem{thas}
T. Hartman and A. Strominger, in progress.

    %\cite{Carlip:1994gc}
\bibitem{Carlip:1994gc}
  S.~Carlip and C.~Teitelboim,
  ``Aspects Of Black Hole Quantum Mechanics And Thermodynamics In
  (2+1)-Dimensions,''
  Phys.\ Rev.\  D {\bf 51}, 622 (1995)
  [arXiv:gr-qc/9405070].
  %%CITATION = PHRVA,D51,622;%%

  %\cite{Maldacena:1998bw}




%%%%%%%%%%%%%%%%%%%%%%%%%%%%%%%%%%%%%%%%%%%






\begin{comment}
NEW REFFERENCES

  %\cite{Banados:2005da}
\bibitem{Banados:2005da}
  M.~Banados, G.~Barnich, G.~Compere and A.~Gomberoff,
  ``Three dimensional origin of Goedel spacetimes and black holes,''
  Phys.\ Rev.\  D {\bf 73}, 044006 (2006)
  [arXiv:hep-th/0512105].
  %%CITATION = PHRVA,D73,044006;%%

  %\cite{Andrade:2005ur}
\bibitem{Andrade:2005ur}
  T.~Andrade, M.~Banados, R.~Benguria and A.~Gomberoff,
  ``The 2+1 charged black hole in topologically massive electrodynamics,''
  Phys.\ Rev.\ Lett.\  {\bf 95}, 021102 (2005)
  [arXiv:hep-th/0503095].
  %%CITATION = PRLTA,95,021102;%%

  %\cite{Witten:1988hc}
\bibitem{Witten:1988hc}
  E.~Witten,
  ``(2+1)-Dimensional Gravity as an Exactly Soluble System,''
  Nucl.\ Phys.\  B {\bf 311}, 46 (1988).
  %%CITATION = NUPHA,B311,46;%%

%\cite{Witten:2007kt}
\bibitem{Witten:2007kt}
  E.~Witten,
  ``Three-Dimensional Gravity Revisited,''
  arXiv:0706.3359 [hep-th].
  %%CITATION = ARXIV:0706.3359;%%

  %\cite{Maloney:2007ud}
\bibitem{Maloney:2007ud}
  A.~Maloney and E.~Witten,
  ``Quantum Gravity Partition Functions in Three Dimensions,''
  arXiv:0712.0155 [hep-th].
  %%CITATION = ARXIV:0712.0155;%%

  %\cite{Yin:2007gv}
\bibitem{Yin:2007gv}
  X.~Yin,
  ``Partition Functions of Three-Dimensional Pure Gravity,''
  arXiv:0710.2129 [hep-th].
  %%CITATION = ARXIV:0710.2129;%%

%\cite{Yin:2007at}
\bibitem{Yin:2007at}
  X.~Yin,
  ``On Non-handlebody Instantons in 3D Gravity,''
  arXiv:0711.2803 [hep-th].
  %%CITATION = ARXIV:0711.2803;%%

\end{comment}

\end{thebibliography}
\end{document}